\documentclass{article}
\usepackage{graphics}
\usepackage{epsf}

\begin{document}
\section*{References for tensile strength of vitreous silica fibers at room
temperature.
\newline \vspace{-23pt}\newline
{\small \textmd{(Compiled by Andri Gretarsson, 7/99)}}}

Many measurements have been reported on the breaking strength of fibers manufactured from 
naturally occurring, and synthetic, vitreous silica. Values reported for the tensile
strength of fibers at room temperature vary greatly, from less than
100 MPa to about 9 GPa. (See Fig. 1.)
The highest strengths are exhibited by untreated
but carefully drawn and carefully handled, 
so-called ``pristine,'' fused silica fibers. With proper manufacture and handling,
it seems possible to obtain extremely high tensile strengths, on the order of several
gigapascals at room temperature, in fibers with diameters as large as 1 mm.
Reduction in the strength is believed to be primarily due to microcracks
induced by normal handling and abrasion rather than by cooling
during manufacture or by aging.


\pagebreak[4]
\begin{figure}
\begin{center}
\epsfxsize=14cm
\leavevmode
\epsfbox{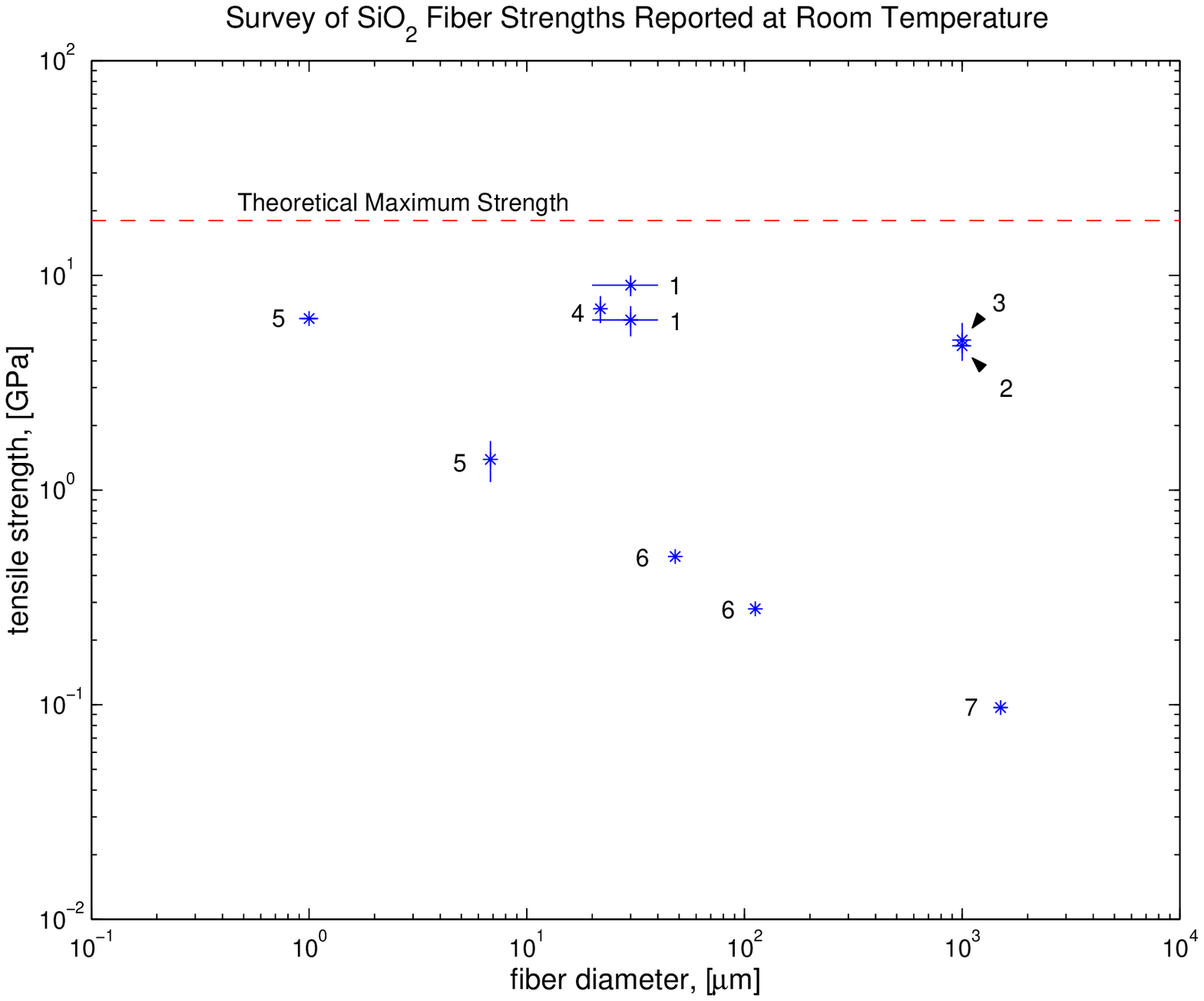}
\end{center}
\caption{Tensile strength reported for vitreous silica fibers (synthetic fused silica
and natural fused quartz) at room temperature.  Numbers correspond to the
references in which the values where reported. Error bars indicate the
uncertainty quoted by the authors where this was available. Otherwise errorbars
represent the last significant figure quoted by the authors or
a range of values given by the authors when they did not provide a precise number.
The theoretical maximum strength is calculated from
the material bond strength, assuming homogeneous stress and
pure ${\mathrm SiO_{2}}$.\cite{Doremus}}
\label{fibstrength}
\end{figure}


\begin{thebibliography}{00}


\bibitem{Proctor} B. A. Proctor, I. Whitney, and J. W. Johnson, Proc. Roy. Soc.
Sec. A, {\bf 297} 534 (1967), and references therein. {\it Twenty-three
page article describing a large number of measurements made by the
authors to test the dependence of the strength of fused silica fibers
and rods on preparation and environmental variables.}

\bibitem{Hillig} W. B. Hillig, J. Appl. Phys. {\bf 32} 741 (1961). {\it
Only one page but informative.}

\bibitem{Morley} J. G. Morley, P. A. Andrews, and J. Whitney, Phys.
Chem. Glasses, {\bf 5} 1 (1964).  

\bibitem{Aslanova} M. S. Aslanova, J. V. Razumovskaya, D. B. Dorzhiev,
and L. A. Sapozhkova, Fiz. Khim. Stekla, {\bf 2} 51 (1976).

\bibitem{Eberhart} E. Eberhart, and H. Kern, H. Klumb, Z. f. angew.
Physik, {3} 209 (1951).

\bibitem{Reinkober} O. Reinkober, Phys. Z. {\bf 33} 32 (1932).

\bibitem{Smekal} A. Smekal, Z. Phys. {\b 114} 448 (1939).

\bibitem{Geppo_private} {\em G. Cagnoli at Glasgow University has recently been making
strength measurements on fused silica and fused quartz
fibers.}\newline


{\bf \large \hspace{-20pt} Reviews, and Books:}

\bibitem{Doremus} R. H. Doremus, {\it Glass Science}, 2$^{\mathrm{nd}}$ ed., John Wiley \& Sons,
New York (1994). {\em Our most used reference book on glass.
Very readable and up to date in its most recent edition.}

\bibitem{Russian_book}  O. V. Mazurin, M. V. Streltsina, and T. P. Shvaiko-Shvaikovskaya,
{\it Handbook of Glass Data, Part A, Silica Glass and Binary Silicate Glasses}, Elsevier
(1983), pp.~117-126. {\em Gives many of the references above. Contains a large
compilation of tables of fiber strength from many sources.}

\bibitem{Ernsberger} F. M. Ernsberger, {\it Strength and Strengthening of
Glass}, in {\it Research Into Glass}, Vol. 2, Glass Research Center, PPG
Industries, Pittsburgh, 1970.  {\it An excellent review article on
the measured strength of glass in general.}

\end{thebibliography}
\end{document}